\documentclass[reprint, amsmath,amssymb,aps, nofootinbib]{revtex4-2}

\usepackage{amssymb}       
\usepackage{amsmath}       
\usepackage{mathtools}     
\usepackage{bm}            
\usepackage{graphicx}      
\usepackage[british]{babel}
\usepackage[T1]{fontenc}
\usepackage[mathlines]{lineno}

\usepackage[table]{xcolor}
\usepackage{multirow}
\usepackage{tabularx}
\usepackage{dcolumn}        

\usepackage[inline]{enumitem} 

\usepackage{newtxtext}                 
\usepackage[slantedGreek]{newtxmath}   

\usepackage{color}                     
\usepackage{soul}                      
\usepackage{hyperref}                   

\usepackage{xifthen}                   
\usepackage{xspace}                    
\usepackage[normalem]{ulem}
\usepackage{booktabs}
\usepackage{lipsum} 

\newcommand{\LCDM}{{$\Lambdaup$CDM}}

\newcommand{\xif}{$\xi(r,a)$~}

\newcommand{\pairvel}{$\text{v}_{12}(r,a)$}
\newcommand{\vot}{$\text{v}_{12}$}  
\newcommand{\orcid}[1]{\href{https://orcid.org/#1}{\,\includegraphics[height=\fontcharht\font`\B]{ORCIDiD.pdf}}}
\newcommand{\github}[1]{\href{https://github.com/#1}{\includegraphics[height=\fontcharht\font`\B]{github-mark.pdf} \nolinkurl{#1}}}

\newcommand{\code}[1]{\texttt{\detokenize{#1}}}

\newcolumntype{L}{>{\raggedright\arraybackslash}X}
\newcolumntype{Y}{>{\centering\arraybackslash}X}
\newcolumntype{R}{>{\raggedleft\arraybackslash}X}

\makeatletter
\newcommand{\phantomlabel}[2]{%
    \protected@write\@auxout{}{%
        \string\newlabel{#2}{{\@currentlabel, #1}{\thepage}{\@currentlabel, #1}{#2}{}}%
    }%
    \hypertarget{#2}{}%
}
\makeatother

\newcommand{\appref}[1]{Appendix~\ref{#1}}
\newcommand{\secref}[1]{Section~\ref{#1}}

\newcommand{\figref}[1]{Fig.~\ref{#1}}

\newcommand{\tabref}[1]{Table~\ref{#1}}

\newcommand{\Quijote}{{\sc Quijote}}
\newcommand{\TNG}{{\sc TNG}}
\newcommand{\TNGtt}{{\sc TNG300-3}}

\newcommand{\astropy}{{\sc astropy}}
\newcommand{\matplotlib}{{\sc matplotlib}}
\newcommand{\numpy}{{\sc numpy}}
\newcommand{\python}{{\sc python}}
\newcommand{\scipy}{{\sc scipy}}

\newcommand{\camb}{{\sc CAMB}}
\newcommand{\pycamb}{{\sc pyCAMB}}

\newcommand{\unit}[1]{\ensuremath{\mathrm{\,#1}}\xspace}
\newcommand{\unitlogicspace}[2]{%
  \ifthenelse{\isempty{#1}}%
    {\unit{#2}}%
    {\ensuremath{{#1}\,\unit{#2}}}%
}

\newcommand{\Mpc}[1][]{\unitlogicspace{#1}{Mpc}}
\newcommand{\Mpch}[1][]{\unitlogicspace{#1}{\mathit{h}^{-1}\Mpc{}}}

\newcommand{\pMpc}[1][]{\unitlogicspace{#1}{Mpc^{-1}}}

\newcommand{\variablelogicspace}[2]{%
  \ifthenelse{\isempty{#2}}%
    {\ensuremath{#1}}%
    {\ensuremath{{#1}={#2}}}%
}

\newcommand{\Ho}[1][]{\variablelogicspace{H_0}{#1}}

\newcommand{\Quijotecosmo}{\ensuremath{
    \Ho[67.11]\pMpc{},\,
    \Omegaup_{\rm m}=0.3175,\,
    \Omegaup_{\Lambda}=0.6825,\,
    \Omegaup_{\rm b}=0.049,\,
    n_{\rm s}=0.9624,\,
    \sigma_8=0.834
}}

\newcommand{\QuijotecosmoNoSigma}{\ensuremath{
    \Ho[67.11]\pMpc{},\,
    \Omegaup_{\rm m}=0.3175,\,
    \Omegaup_{\Lambda}=0.6825,\,
    \Omegaup_{\rm b}=0.049,\,
    n_{\rm s}=0.9624
}}

\newcommand{\TNGcosmo}{\ensuremath{
    \Ho[67.74]\pMpc{},\,
    \Omegaup_{\rm m}=0.3089,\,
    \Omegaup_{\Lambda}=0.6911,\,
    \Omegaup_{\rm b}=0.0486,\,
    n_{\rm s}=0.9667,\,
    \sigma_8=0.8159
}}

\graphicspath{{Figures/}}

\hypersetup{
    colorlinks = true,
    linkcolor = blue,
    citecolor = blue,
    urlcolor  = blue
}

\begin{document}
\preprint{APS/123-QED}
\title{Non-linear Pairwise Velocities as a Cosmological Probe}
\author{Mariana Jaber}%
\email{ \{ mariana.jaber, maria.taverna, rasmus.strid \}@ncbj.gov.pl}
\author{Antonela Taverna}
\author{Rasmus Strid}
\affiliation{National Centre for Nuclear Research, Pasteura 7, 02-093 Warsaw, Poland}
\author{Wojciech A. Hellwing}
\affiliation{Center for Theoretical Physics, Polish Academy of Sciences, Al. Lotnik\'ow 32/46, 02-668 Warsaw, Poland }

\date{\today}

\begin{abstract}

Peculiar velocities trace gravitational dynamics directly, complementing density-based clustering as wide-field spectroscopic surveys (DESI, Euclid, PFS) and kinetic Sunyaev--Zel'dovich (kSZ) measurements enter the precision era. Observational analyses of pairwise velocities and the pairwise kSZ signal remain largely restricted to linear and quasi-linear scales ($\gtrsim20$--$30\,\Mpch$), despite substantial cosmological information available at smaller separations.

We present a simulation-calibrated likelihood for \pairvel, built on the exact pair conservation equation and \pycamb\ HMcode-2020 non-linear clustering, and validate it against the \Quijote\ and \TNGtt\ suites across resolution, particle sampling, box size, and redshift.

Using \Quijote, we show that extending the fit to non-linear scales tightens the $1\sigma$ uncertainty on $\Omega_\mathrm{m}$ and $\sigma_8$ by $74\%$ and $81\%$ at $z=0.5$ ($r_{\rm min}=4\,\Mpch$), and by $70\%$ and $74\%$ at $z=0$ ($r_{\rm min}=8\,\Mpch$), relative to the linear-regime baseline ($r_{\rm min}=50\,\Mpch$) (all at fixed $r_{max}=140\,\Mpch$); $\sigma(f\sigma_8)$ tightens by up to $86\%$, and the joint $\Omega_\mathrm{m}$--$f\sigma_8$ Figure-of-Merit improves by up to $9.1\times$ relative to the same baseline. These constraints are conditional on $h$ fixed at its \Quijote\ fiducial value, due to a near-degenerate response with $\sigma_8$; jointly sampling $h$ shifts and widens them substantially. This framework provides a validated route to exploiting that information in upcoming direct peculiar-velocity, redshift-space, and kSZ analyses.

\end{abstract}

\maketitle


\section{\label{sec:intro}Introduction}
The large-scale structure of the Universe traces its origin to gravitational amplification of primordial density fluctuations. While galaxy clustering remains foundational, peculiar velocities provide complementary information about the gravitational dynamics of structure formation.
The mean pairwise velocity \pairvel, defined as the pair-weighted ensemble-averaged relative radial velocity of pairs separated by proper distance $r$, can be derived from the Bogoliubov--Born--Green--Kirkwood--Yvon (BBGKY) hierarchy.

The pair conservation equation links pairwise motion to the two point correlation function and its time evolution \citep{davisIntegrationBBGKYEquations1977, peeblesLargescaleStructureUniverse1980, juszkiewiczSkewedExponentialPairwise1998}, exhibiting distinct non-linear (virialised) and linear (growing mode) limits.
In this work, we model and measure \pairvel\ for dark-matter particles; its application to observed galaxy or halo samples additionally requires accounting for tracer selection and pair weighting.

Peculiar-velocity surveys have progressively constrained the growth combination $f\sigma_8$  at $z\sim 0$, converging to $f\sigma_8\approx0.310$-- $0.471$ \citep{turnerDESIDR1Peculiar2025} in recent compilations. 
However these measurements remain limited by the intrinsic $f$--$\sigma_8$ degeneracy, cosmic variance arising from sparse and anisotropic sampling, and systematic uncertainties associated with our location in a structured environment (e.g. proximity to the Virgo cluster, \cite{hellwingNotCopernicanObserver2017}). These effects complicate unbiased inference of growth parameters and limit the statistical robustness of present constraints.

Modern datasets, including DESI peculiar velocity measurements and pairwise kSZ detection from CMB surveys such as ACT and Planck \citep{ hadzhiyskaProbingCosmicVelocities2025, gongDetectionPairwiseKinematic2025} have achieved high-significance measurements of velocity statistics. 
On the direct peculiar-velocity side, the forthcoming 4MOST surveys, in particular the 4MOST Hemisphere Survey (4HS), will substantially enlarge the available low-redshift velocity samples \citep{dejong4MOST4metreMultiobject2012,taylor4MOSTHemisphereSurvey2023}.
These analysis typically interpret the observed pairwise kSZ signal by comparison to linear or quasi-linear theory predictions on scales above $\sim 20-30$\Mpch, where such approximations are adequate.  However, as survey precision improves, robust modelling beyond the strictly linear regime -- including validated non-linear behaviour and realistic covariance -- becomes increasingly important for unbiased cosmology inference.

Analytical and simulation-based modelling of pairwise velocities beyond the strictly linear regime has a long history \citep{juszkiewiczDynamicsPairwiseMotions1999}, with recent work extending and testing such approaches deep into the non-linear regime \citep{jaberDynamicsPairwiseMotions2023,maleubreConstrainingAccuracyPairwise2023}.
We present a simulation calibrated likelihood for \pairvel~ based on the exact pair conservation equation \citep{juszkiewiczSkewedExponentialPairwise1998, jaberDynamicsPairwiseMotions2023}, using the \Quijote~simulations \cite{villaescusa-navarroQuijoteSimulations2020} across variations in resolution, particle subsampling, and redshift.

This framework provides a robust route to extracting growth constraints from pairwise velocity measurements in the era of precision large-scale structure surveys.

Section~\ref{sec:methods} details our model testing and implementation. Section~\ref{sec:model} introduces the model, Section~\ref{sec:implementation} describes the numerical implementation and the suite of simulations used, we describe our different tests in Section~\ref{sec:convergence}, and our statistical treatment in Section~\ref{sec:likelihood}. Section~\ref{sec:results_data} validates the model against simulations and establishes the resolved scale range, and Section~\ref{sec:results_validation} presents systematic tests. Section~\ref{sec:sensitivity} quantifies the sensitivity of \pairvel~to $h$, $\Omega_\mathrm{m}$, and $\sigma_8$, and Section~\ref{sec:constraints} presents stand-alone MCMC constraints on $\Omega_\mathrm{m}$, $\sigma_8$, $f\sigma_8$ and $f$. We discuss the implications for cosmological constraints from pairwise motions in Section~\ref{sec:discussion}, and present our concluding remarks in Section~\ref{sec:conclusions}.

\section{\label{sec:methods}Methods}
We implement the exact pair conservation equation to predict the mean pairwise velocity \pairvel~ and construct a likelihood for comparison with $N$-body simulations. Unlike a simple model implementation, our focus is on quantifying the numerical stability of cosmological constraints under variations in binning, derivative evaluation, Fourier range, simulation volume, redshift, and non-linear power spectrum prescription. Establishing convergence across these choices is essential for robust inference from pairwise velocity statistics.

\subsection{Pairwise velocity model}\label{sec:model}

The mean pairwise velocity \pairvel~ follows from the exact pair conservation equation derived from the BBKGY hierarchy \citep{juszkiewiczDynamicsPairwiseMotions1999, juszkiewiczSkewedExponentialPairwise1998}.
In this framework, the velocity is fully determined by the two-point correlation function, $\xi(r,a)$, and its time evolution.
Accurate prediction of \pairvel~ therefore, reduces to stable computation of $\xi(r,a)$, its spherical average, $\bar{\xi}(r,a)$, and its time evolution, $\partial_a\bar{\xi}(r,a)$.

The starting equation for our model is:
\begin{equation}
\label{eq:v12}
\text{v}_{12}(r,a)=-\frac{H(a)ra^2\partial_a\bar{\xi}(r,a)}{3[1+\xi(r,a)]},
\end{equation}
where $r$ is the comoving separation\footnote{Unlike in \cite{juszkiewiczDynamicsPairwiseMotions1999}, we use \emph{r} to refer to co-moving separations throughout the paper.} of the pair, $\xi(r,a)$ is the two point correlation function  and $\bar{\xi}(r,a) \equiv 3r^{-3}\int_0^r\xi(s,a)s^2\,ds$, is its spherical volume average.

In the non-linear regime ($r\ll1$, $\xi\gg1$), close pairs reside in virialised haloes whose random, isotropic orbital motions saturate the dynamics, so their net streaming along the line of centres must vanish as $r\rightarrow 0$; these random virial motions instead contribute primarily to the pairwise velocity dispersion.

At large scales, $\xi(r,a)\approx D(a)^2\xi_0(r)$ recovers:
\begin{equation}\label{eq:v12lin}
\text{v}_{12}(r,a)=-\frac{2}{3}r a H f \bar{\bar{\xi}}_{\text{0}}(r,a),
\end{equation}
where $\bar{\bar{\xi}}\equiv\bar{\xi}/[1+\xi]$, $\xi_0$ is the linear correlation function, and $f=d\ln D/d\ln a$. On scales of $r\approx 50\;h^{-1}\;\text{Mpc}$, the linear approximation (Eq.~\ref{eq:v12lin}) agrees with the full non-linear solution (Eq.~\ref{eq:v12}) within $5\%$. On smaller scales however, Eq.~\ref{eq:v12lin} underestimates the amplitude of infall velocities and so the accurate evaluation of \(\xi(r,a)\), \(\bar{\xi}(r,a)\), and \(\partial_a \bar{\xi}(r,a)\) is therefore the central task for robust cosmological inference.

\subsection{Numerical implementation}\label{sec:implementation}
The theoretical prediction of \pairvel~ requires three quantities computed from the matter power spectrum: the two-point correlation function $\xi(r,a)$, its spherical average $\bar{\xi}(r,a)$, and its scale-factor derivative $\partial_a\bar{\xi}(r,a)$. We obtain the non-linear matter power spectrum $P_\mathrm{nl}(k,a)$ using \code{HMcode-2020} as implemented in \pycamb~\citep{meadHMcode2020ImprovedModelling2021}, and transform to
configuration space via the Fourier--Bessel integral:
\begin{equation}
\xi(r,a) = \frac{1}{2\pi^2}\int_0^\infty k^2 P_\mathrm{nl}(k,a)\,
j_0(kr)\,dk,
\end{equation}
with $j_0(kr)=\sin(kr)/(kr)$.The spherical average $\bar{\xi}(r,a)$ and the derivative $\partial_a\bar{\xi}(r,a)$ are then evaluated on logarithmic $r$-bins spanning $0.1$--$140\,\Mpch$.

The Fourier $k$-range is chosen to
encompass all modes supported by the simulation box, ensuring large-scale
power is accurately captured. The scale-factor derivative is approximated
using central differences across a set of \pycamb~snapshots; the
sensitivity to the number and spacing of these snapshots is explicitly
tested in Section~\ref{sec:results_validation}.

The simulation-measured \pairvel~ defines the data vector used in the likelihood.  
We compute \pairvel~ from the suite of $N$-body simulations listed in Table~\ref{tab:sims}. 
The \Quijote~ runs use the Planck~2018 \citep{aghanimPlanck2018Results2018} fiducial cosmology (\Quijotecosmo), while \TNGtt~ adopts the Planck~2015~\citep{adePlanck2015Results2016} values (\TNGcosmo). 
The production analysis uses \Quijote--Mid-Resolution snapshots at $z=0$ and $z=0.5$ with 10\% particle thinning.
For validation, we also use the full Mid-Resolution snapshots at $z=0$, \Quijote--High-Resolution runs at $z=0$ with 10\% thinning, and \TNGtt~ at $z=0$ with 10\% thinning, enabling assessment of convergence with respect to particle sampling, box size, and resolution.

\begin{table}
\centering
\small
\setlength{\tabcolsep}{4pt}
\begin{tabular}{lccccc}
\hline
Name & Box & $N_\mathrm{tr}$ & $z$ & Runs & Thin. \\
 & [$h^{-1}$Mpc] &  &  &  &  \\
\hline
\textbf{QUIJOTE MR} & \textbf{1000} & \textbf{$237^3$} & \textbf{0, 0.5} & \textbf{100} & \textbf{10\%} \\
\hline
QUIJOTE MR & 1000 & $512^3$ & 0, 0.5 & 100 & 100\% \\
QUIJOTE HR & 1000 & $475^3$ & 0 & 50 & 10\% \\
\TNGtt~ LR & 205 & $290^3$ & 0 & 1 & 10\% \\
\hline
\end{tabular}
\caption{
\textbf{Simulation suite used for \pairvel\ measurements.}
Production analyses use \Quijote~ mid-resolution (MR) simulations with 10\% thinning at $z=0$ and $z=0.5$.
Validation tests probe thinning, resolution convergence, and box-size dependence. All \Quijote~ simulations assume the Planck 2018 cosmology \cite{aghanimPlanck2018Results2018}, while \TNGtt~ adopts the Planck 2015 cosmology \cite{adePlanck2015Results2016}. 
}
\label{tab:sims}
\end{table}

\subsection{Robustness and validation strategy}\label{sec:convergence}
Before using the likelihood for cosmological inference, we establish that the model predictions and resulting parameter constraints are stable under the numerical choices inherent to our pipeline. These tests fall into two complementary categories.

The first concerns the robustness of the \textit{theoretical modelling} itself, independent of any particular simulation. We compare several non-linear power spectrum prescriptions available in \pycamb, finding that \code{HMcode-2020} uniquely stabilizes the scale-factor derivative $\partial_a\bar{\xi}(r,a)$ across all redshifts tested; other prescriptions  introduce numerical artefacts in the derivative that propagate into \pairvel. Within the chosen prescription, we verify that the results are insensitive to the discretization of the redshift-array used for the central-difference derivative and to the Fourier $k$-range used in the Bessel transform.

The second category concerns simulation systematics: whether the measured \pairvel~ depends on particle thinning fraction, simulation resolution, box size, or redshift. These tests are carried out across the full validation suite in Table~\ref{tab:sims}, with \TNGtt~ providing an independent cross-check at a different box size and cosmology. 
Detailed results for all tests are presented in \secref{sec:results}.

\subsection{Likelihood and Covariance}\label{sec:likelihood}

Because the number of fitted radial bins is a non-negligible fraction of the number of independent realisations available to estimate the covariance (up to $p=20$ bins from $N=100$ realisations), a Gaussian likelihood built from the raw sample covariance underestimates parameter uncertainty: the inverse covariance $C^{-1}$ is not only biased in expectation but is itself a noisy, Wishart-distributed quantity, and a single point estimate does not capture that sampling uncertainty. We therefore adopt the multivariate $t$-distribution likelihood of \citet{sellentinParameterInferenceEstimated2016}, which marginalises analytically over the covariance's own sampling distribution:
\begin{equation}
-2\ln\mathcal{L}(\boldsymbol{\theta}) = N\ln\!\left[1+\frac{\chi^2(\boldsymbol{\theta})}{N-1}\right],
\end{equation}
\begin{equation}
\chi^2(\boldsymbol{\theta}) = [\hat{\mathbf{v}}_{12} - \mathbf{v}_{12}(\boldsymbol{\theta})]^\top
C^{-1} [\hat{\mathbf{v}}_{12} - \mathbf{v}_{12}(\boldsymbol{\theta})],
\end{equation}
using the sample covariance $C^{-1}$ directly, without the finite-sample debiasing rescaling $C^{-1}\to\alpha C^{-1}$ (with $\alpha=(N-p-2)/(N-1)$) that corrects only the estimator's expectation value while leaving its own sampling uncertainty unaccounted for \citep{Hartlap2007}.  $\hat{\mathbf{v}}_{12}$ is the per-bin mean of the \Quijote~measurements across the radial bins defined by the scale range under consideration, taken over the $N=100$ independent realisations; $\mathbf{v}_{12}(\boldsymbol{\theta})$ is the model prediction, and $C$ is the bin-to-bin covariance estimated from the same $N=100$ independent realizations via the unbiased sample estimator,
\begin{equation}
\label{eq:covmat}
C_{ij} = \frac{1}{N-1} \sum_{r=1}^{N} 
\left(x_i^{(r)} - \bar{x}_i\right)\left(x_j^{(r)} - \bar{x}_j\right),
\end{equation}
where $x_i^{(r)}$ is the value of $\text{v}_{12}$ in bin $i$ for realisation $r$
and $\bar{x}_i$ is the mean over realisations.

Posterior sampling is carried out using a Metropolis--Hastings Markov Chain Monte Carlo (MCMC) algorithm implemented in \code{Cobaya} \cite{torradoCobayaCodeBayesian2021}, the same sampler used in \code{CosmoMC} and subsequently ported to \code{Cobaya}, varying $\Omega_\mathrm{c} h^2\in[0.05,0.25]$ and $\ln(10^{10}A_s)\in[2,4]$ under uniform priors, with $h$ fixed at the \Quijote~fiducial value ($h_\mathrm{fid}=0.6711$) and all remaining cosmological parameters fixed to the \Quijote~fiducial cosmology (\QuijotecosmoNoSigma). We do not sample $h$ freely alongside these parameters: as shown in \secref{sec:sensitivity} (\figref{fig:sensitivity}), \pairvel~responds to $h$ and $\sigma_8$ with similar scale dependence over the fitted range, so pairwise-velocity data alone cannot separate the two.
In addition, we examine the posterior distributions of the derived parameters $\Omega_m, \sigma_8, f\sigma_8$, and $f$.
The main production analyses employ the Mid-Resolution 10\% covariance matrix calculated using 100 independent realisations, whereas robustness is assessed by means of validation runs using both the full Mid-Resolution (100\%) and \Quijote--High-Resolution (10\%) simulations. Convergence of the MCMC chains is evaluated using the Gelman–Rubin diagnostic \cite{lewisEfficientSamplingFast2013}, requiring $R-1 < 0.01$, with an initial burn-in phase corresponding to $50\%$ of the total samples.

\section{\label{sec:results}Results}
\subsection{Simulation validation and resolved scales}\label{sec:results_data}
\begin{figure}
    \centering
    \includegraphics[width=\columnwidth]{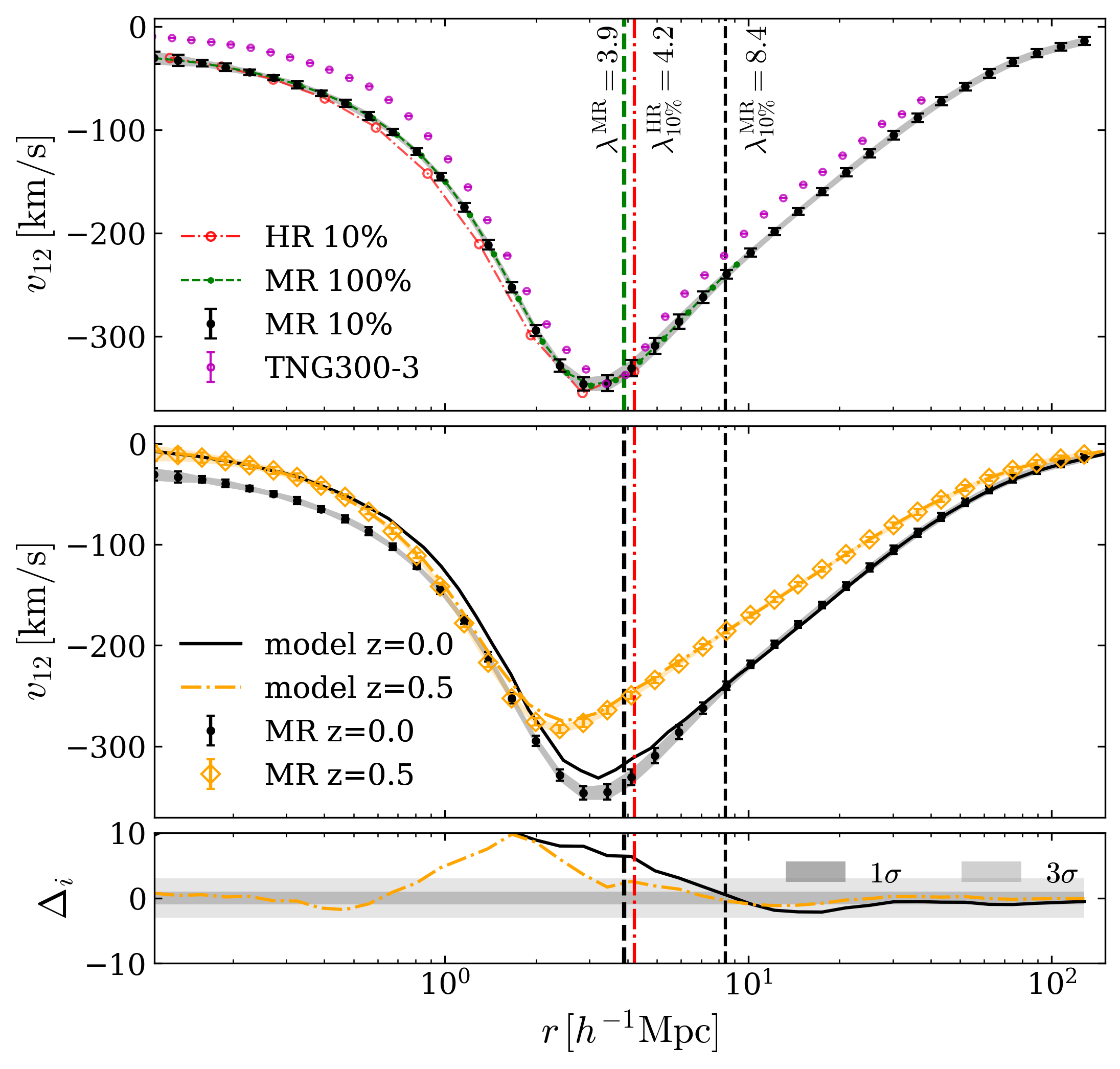}
    \caption{
        \textbf{Simulation validation of the pairwise velocity model.}
    \textbf{Top}: Convergence of the \Quijote~ \pairvel~ measurement at $z=0$ across particle sampling and mass resolution. Mid-Resolution measurements at $100\%$ and $10\%$ (production) thinning agree above their respective Nyquist scales, $\lambda_{\rm Nyq}^{\rm MR}=3.9\,h^{-1}$Mpc and $\lambda_{\rm Nyq,10\%}^{\rm MR}=8.4\,h^{-1}$Mpc, as do the independent High-Resolution ($10\%$) and TNG300-3 measurements, validating the use of the Mid-Resolution $10\%$ configuration as the production dataset.
    \textbf{Middle}: Mean pairwise velocity \vot from \Quijote~ Mid-Resolution (10$\%$ thinning, production configuration) measurements at $z=0$ and $z=0.5$, compared to the theoretical prediction (solid/dash-dotted lines). Shaded bands show $\pm\sqrt{C_{ii}}$, the square root of the diagonal covariance from $N=100$ independent realizations.
    \textbf{Bottom:} Bin-wise normalized residuals, $\Delta_i \equiv (v^{\rm model}_i - v^{\rm data}_i)/\sqrt{C_{ii}}$, with shaded $\pm1\sigma$ and $\pm3\sigma$ regions. At both redshifts, $|\Delta_i|$ exceeds $3\sigma$ at small separations ($r\lesssim6\,h^{-1}$Mpc at $z=0$; $r\approx0.9$--$3\,h^{-1}$Mpc at $z=0.5$) and settles within $\sim1\sigma$ only near or beyond the respective Nyquist scale, motivating the scale cuts adopted for parameter inference (\secref{sec:constraints}).}
    \label{fig:v12model}
\end{figure}

Figure \ref{fig:v12model} establishes the scale range over which \Quijote~\vot~measurements are reliable, by systematically testing sensitivity to particle thinning and mass resolution. These tests set the minimum usable scale, $r_{\rm min}$: at $z=0$, the tested \Quijote~configurations show a small-scale loss of model--data agreement below their respective resolved scales, a small-scale limitation that leads our production inference to emphasize the $z=0.5$ snapshot, retaining $z=0$ only for completeness (\secref{sec:constraints}). The maximum usable scale, $r_{\rm max}$, is in principle bounded by the simulation box size through its fundamental mode; we validate in \appref{app:lbox} that this does not bias our production choice of $r_{\rm max}=140\,\Mpch$ for the large \Quijote~volume, though it is a more binding constraint for smaller-volume simulations such as TNG300-3.

At $z=0$, the Mid- and High-Resolution \Quijote~ configurations agree above their respective Nyquist scales, demonstrating convergence with respect to particle sampling and resolution (top panel, Fig.~\ref{fig:v12model}). This validates the use of the Mid-Resolution 10\% thinning configuration as the production dataset.

Two complementary comparisons help assess the small-scale behaviour identified above.
First, TNG300-3, which resolves smaller scales, provides an independent higher-resolution cross-check of the small-scale \pairvel~measurement (top panel, Fig.~\ref{fig:v12model}); we note that TNG300-3 contributes a single realisation with no covariance estimate, and therefore serves as a qualitative cross-check only.
Second, the \Quijote~ Mid-Resolution measurements at $z=0.5$ show controlled model--data residuals down to smaller separations. Together with the thinning and resolution tests, these comparisons motivate a conservative exclusion of scales at which the \pairvel~measurement or model--data residuals are not demonstrably converged.

Based on this residuals test, we adopt $r_{\rm min}=8\,h^{-1}$Mpc at $z=0$, matching the production Nyquist scale. At $z=0.5$ the model and data already agree within $3\sigma$ down to $r\approx3\,h^{-1}$Mpc, supporting the smaller $r_{\rm min}=4\,h^{-1}$Mpc adopted there (\secref{sec:constraints}).

\subsection{Model robustness across specifications}\label{sec:results_validation}

Before assessing how non-linear scales sharpen cosmological constraints relative to those from the linear streaming regime (i.e. separations where the linear theory approximation of (Eq. \eqref{eq:v12lin}) matches the full non-linear model (Eq. \eqref{eq:v12}) to within $5\%$ (\secref{sec:constraints})) we must first verify the reliability of the numerical method used to solve for \pairvel~. Building on the simulation–measurement validation in \secref{sec:results_data}, we test the robustness of the theoretical modelling: how the predicted \pairvel~ depends on the non-linear matter power-spectrum prescription and on the redshift-array discretization used to evaluate $\partial_a\bar{\xi}(r,a)$.

\begin{figure}
\includegraphics[width=\columnwidth]{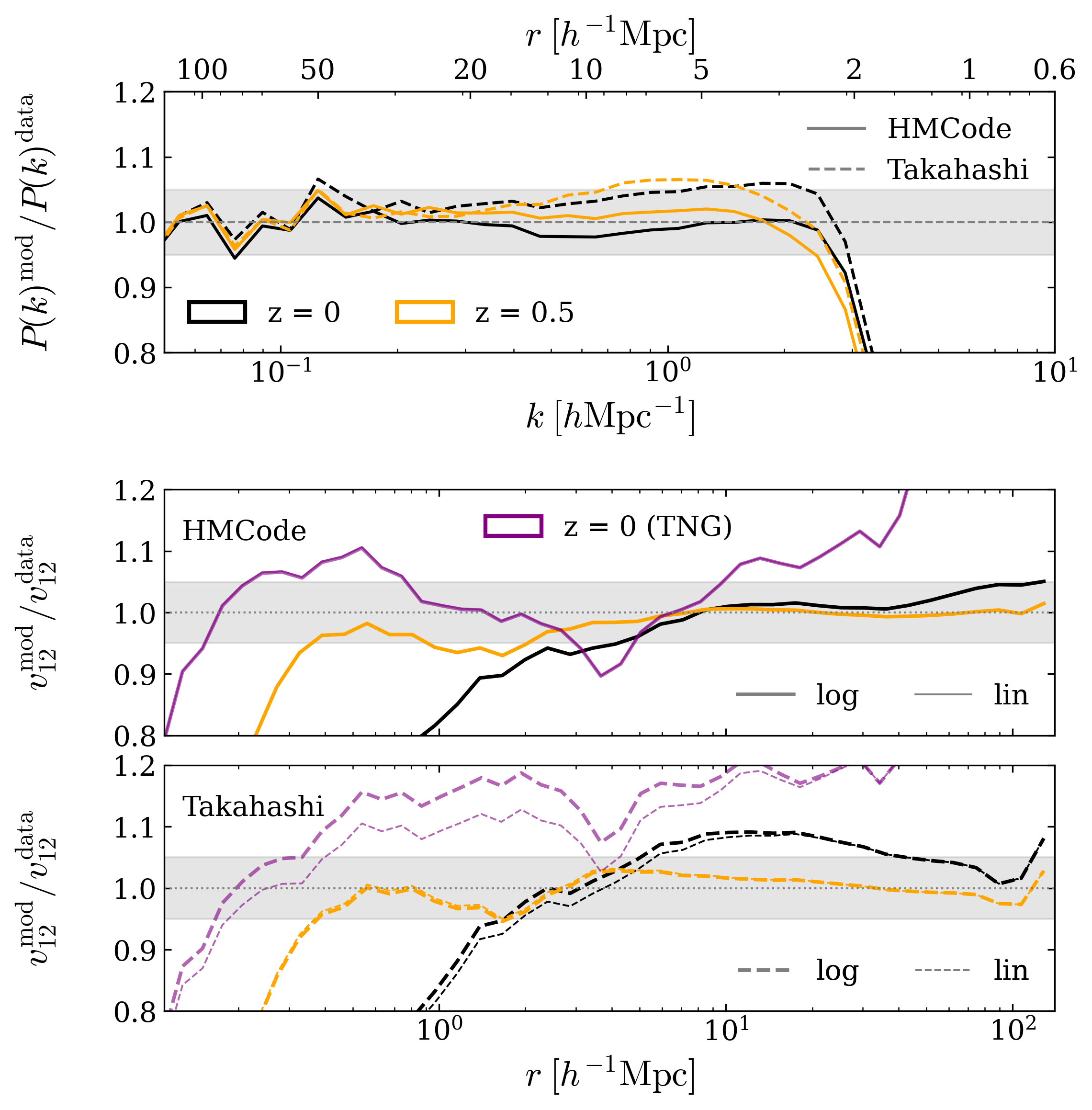}
\caption{
\label{fig:halofit}
\textbf{Sensitivity of the \pairvel\ model to the non-linear power-spectrum
prescription and to the discretization of the redshift array used in the model
evaluation.}
\textbf{Top:} Ratio of the modelled non-linear matter power spectrum to the
\Quijote\ measurement, $P(k)^{\rm mod}/P(k)^{\rm data}$, at $z=0$ and
$z=0.5$, computed with \code{HMcode-2020} (solid) and the Takahashi
\citep{takahashiRevisitingNonlinearMatter2012} \textsc{halofit} prescription
(dashed). The upper axis shows the corresponding scale $r=2\pi/k$.
\textbf{Bottom panels:} Ratio $v_{12}^{\rm mod}/v_{12}^{\rm data}$ for the \code{HMcode-2020} (middle) and Takahashi (lower) prescriptions. Each panel shows $z=0$ and $z=0.5$ models, divided by the \Quijote\ data, together with $z=0$ from \TNG\, divided by the \TNG\ measurement. Thick lines use a logarithmically spaced redshift array in the model evaluation; thin lines use a linearly spaced one. \code{HMcode-2020} yields predictions that are essentially independent of this choice, whereas the Takahashi prescription shows substantially stronger sensitivity to the redshift discretization. The shaded bands represent $\pm 5\%$ variations.}
\end{figure}

Figure~\ref{fig:halofit} presents the results of this comparison.
To ensure the robustness of these conclusions, we additionally perform an extended validation using the \TNG300-3 simulation, providing an independent assessment of the model's numerical stability beyond the primary Quijote suite.

The top panel shows the ratio of the modelled non-linear matter power spectrum to the \Quijote~ measurement, $P(k)^{\rm mod}/P(k)^{\rm data}$, at $z=0$ and $z=0.5$, for the \code{HMcode-2020}~\citep{meadHMcode2020ImprovedModelling2021} and Takahashi~\citep{takahashiRevisitingNonlinearMatter2012} prescriptions.
Over the range of scales relevant to \pairvel, \code{HMcode-2020} provides the more accurate description of the non-linear power spectrum.
Using the Takahashi prescription, previously adopted for \vot~modelling in \citet{jaberDynamicsPairwiseMotions2023}, the predicted \pairvel~ shows a strong dependence on redshift discretization at all redshifts. In contrast, \code{HMcode-2020} gives stable predictions at all tested redshifts ($z=0$ and $0.5$), with negligible variation between redshift arrays.

We therefore adopt HMcode-2020 as the fiducial non-linear prescription throughout this work, a choice further supported by its closer agreement with the measured $P(k)$ in the top panel. With this choice, derivative-discretization effects are negligible at the redshifts used for parameter inference.

Previous studies have reported box-size effects in cosmological N-body simulations, such as the finite-volume suppression of halo velocity correlations and the resulting biases in inferred parameters like $f\sigma_8$ \cite{chuangLargescaleHaloVelocity2026}. Consequently, given that we employ the TNG-300 simulation to validate our findings, we demonstrate in \appref{app:lbox} that our conclusions are robust to the finite volume of the simulation box.

Figure~\ref{fig:halofit} also shows that the model-data agreement is tighter for \Quijote~ than for \TNG: using \code{HMcode-2020}, the \Quijote~ ratios remain within or close to the shaded $\pm5\%$ band across most of the scales shown, while the \TNG~ ratio departs more substantially, exceeding $10\%$ over an extended range of intermediate-to-large scales. Together with the larger number of independent realizations available for covariance estimation, this motivates using \Quijote~ as the primary suite for parameter inference throughout this work, despite the small-scale $z=0$ mismatch discussed in \secref{sec:results_data}; \TNG~ instead serves as a targeted cross-check (e.g.\ \appref{app:lbox}).

\subsection{Cosmological sensitivity}\label{sec:sensitivity}

\begin{figure}
    \includegraphics[width=\columnwidth]{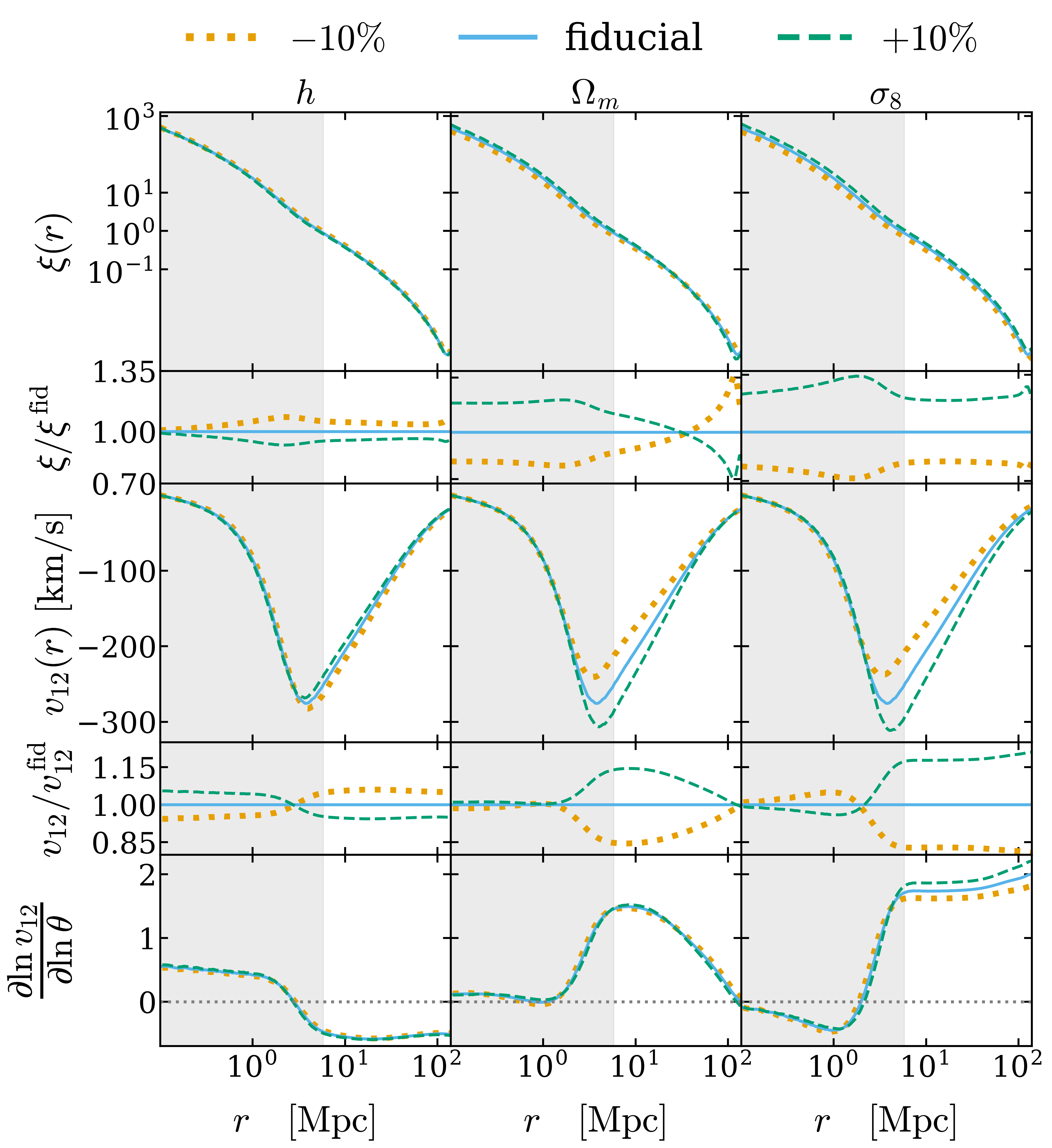}
    
    \caption{\textbf{Sensitivity of $\xi(r,a)$ and $\text{v}_{12}(r,a)$ to cosmological parameters.} Each column varies one parameter independently:  $h$ (left), $\Omega_m$ (centre), $\sigma_8$ (right), by $10\%$ around the fiducial \Quijote~ cosmology (solid, light blue), with the colour scale running from minimum (dotted, orange) to maximum (dashed, green) parameter value. \textbf{Rows 1--2:} Two-point correlation function $\xi(r,a)$ and its ratio to the fiducial, $\xi/\xi^{\rm fid}$. \textbf{Rows 3--4:} Mean pairwise velocity \pairvel~ and its ratio to the fiducial, $\text{v}_{12}/\text{v}_{12}^{\rm fid}$. \textbf{Row 5:} Logarithmic derivative $\partial\ln \text{v}_{12}/\partial\ln\theta$, quantifying the scale-dependent response of \pairvel~ to each parameter. Every quantity is evaluated at $z = 0.5$. The shaded grey band indicates scales below the Nyquist frequency of the production dataset, $r \leq \lambda^{\text{MR}}_{\text{Nyq}}$.
\label{fig:sensitivity}}
\end{figure}

Having disentangled the effects of simulation choices (\secref{sec:results_data}) and of the non-linear numerical model (\secref{sec:results_validation}) from the underlying signal, we now isolate the response of \pairvel~ to changes in cosmology alone.

Figure~\ref{fig:sensitivity} shows the variation of
$\xi(r,a)$ and $\text{v}_{12}(r,a)$ as each parameter $\theta=\{\,h,\,\Omega_m,\sigma_8\}$ is varied independently around the fiducial \Quijote~ cosmology, alongside the logarithmic derivative $\partial\ln\text{v}_{12}/\partial\ln\theta$. When varying $\Omega_m$, the physical baryon density parameter $\omega_b=\Omega_bh^2$ is kept fixed, so a variation in $\Omega_m$ is directly proportional to a variation in $\omega_c=\Omega_ch^2$. 

For variations in $h$, the physical density parameters $\omega_b,\;\omega_c$ are kept fixed and separations are expressed in units of $[\text{Mpc}]$ rather than $[h^{-1}\text{Mpc}]$, to separate changes in the signal shape from unit conversion from physical effects of changing the cosmology.

We quantify the scale-dependent response to parameter variations via the logarithmic derivatives $\partial\ln\text{v}_{12}/\partial\ln\theta$ shown in the bottom row of \figref{fig:sensitivity}. 
While \xif~shows largest variation on smaller scales, slowly decreasing around intermediate scales, $r\sim 10$ Mpc, \pairvel~ shows a distinctly different trend, with variation increasing in size at intermediate scales for all parameters.\\

The $\sigma_8$ response is the most uniform across intermediate and larger scales, $\partial\ln\text{v}_{12}/\partial\ln\sigma_8\approx1.6$ at intermediate separations ($5$--$20\,\text{Mpc}$), rising gradually to $\approx1.9$ at larger separations ($r\gtrsim50\,\text{Mpc}$), consistent with the near-linear-theory expectation of $2$ and reflecting a near-pure amplitude rescaling of the clustering signal. This limiting value has a simple analytic origin in Eq. \eqref{eq:v12lin}. Since $\bar{\bar\xi}=\bar\xi/(1+\xi)$ and the linear correlation function scales as $\xi\propto\sigma_8^2$ at fixed shape, differentiating the full ratio with respect to $\sigma_8$ gives $\partial\ln\text{v}_{12}/\partial\ln\sigma_8=2/[1+\xi(r)]$, an exact relation within the linear approximation of Eq. \eqref{eq:v12lin} \citep{juszkiewiczDynamicsPairwiseMotions1999,ferreiraStreamingVelocitiesDynamical1999}. The response depends on scale only through $\xi(r)$ itself: it approaches $2$ smoothly as $\xi(r)\rightarrow0$ at large $r$, and is suppressed below $2$ wherever $\xi(r)$ is not negligible, consistent with the $\approx1.6$ response measured at intermediate scales.\\

Similarly, the Hubble parameter response is near constant for intermediate and large $r$, with $\partial\ln\text{v}_{12}/\partial\ln h \approx -0.57$ on scales of $10-100\;\text{Mpc}$ and likewise on small scales, where $\partial\ln\text{v}_{12}/\partial\ln h \approx 0.45-0.55$ for $r$ below $1\;\text{Mpc}$. In addition, $\partial\ln\text{v}_{12}/\partial\ln h$ changes sign around $r\approx 4-5$, on the same scales as the $\sigma_8$ response.

The response of $\Omega_\mathrm{m}$ has a distinctly different behaviour: it is positive ($\partial\ln\text{v}_{12}/\partial\ln\Omega_\mathrm{m}\approx0.14$) and increasing at intermediate scales, where the shape of the matter transfer function dominates, but starts decreasing at $r\approx20$--$25\,\text{Mpc}$, and drops off at larger scales where the enhanced expansion rate suppresses the normalised infall signal, becoming negative at scales $r\approx 100-110\; \text{Mpc}$. 
The sign reversal on large scales for $\Omega_m$ has no counterpart in the $h,\sigma_8$ responses, nor in \xif, making it a distinctive signature of pairwise velocities.\\

The similarity in the response to $h,\sigma_8$ indicates that the two parameters cannot be constrained by \pairvel~alone within the scales accessed in this work; we therefore fix $h$ at its fiducial value in the joint constraints of \secref{sec:constraints} below, rather than sample it jointly with $\sigma_8$.
In order to jointly sample all three parameters using \pairvel alone, one would need to access the scales where the slope of the responses are sufficiently distinct, around $r\simeq 1\;\text{Mpc}$.\\

The complementarity of these scale-dependent responses shows that \pairvel~ provides parameter sensitivity complementary to that of the equal-time two-point correlation function,
motivating its use across a broad $r$-range for the joint MCMC constraints presented in Section~\ref{sec:constraints}. 

\subsection{Parameter constraints}\label{sec:constraints}

In this section, we present the marginalized posterior constraints on the physically intuitive parameters $\sigma_8$, $\Omega_\mathrm{m}$, and $f\sigma_8$ ($h$ is fixed at its \Quijote~fiducial value throughout, \secref{sec:likelihood}, motivated by the $h$--$\sigma_8$ degeneracy of \figref{fig:sensitivity}), rather than the full set of parameters sampled directly by the MCMC, $\ln(10^{10}A_s)$ and $\Omega_\mathrm{c} h^2$ (\secref{sec:likelihood}); $\sigma_8$, $\Omega_\mathrm{m}$, and $f\sigma_8$ are derived from these within the assumed $\Lambda$CDM model. These constraints are therefore conditional on the fixed fiducial $h$; jointly sampling $h$ shifts and widens them substantially, as quantified in \secref{sec:discussion_degeneracy}, given the degeneracy noted above. The full set, including the sampled parameters, is tabulated in Table~\ref{tab:mcmc_constraints}, Appendix~\ref{app:mcmc_stats}.

We split the constraints by redshift, first discussing the results from $z=0.5$ and then from $z=0$, since the \Quijote~ measurements at $z=0$ do not recover the expected $\text{v}_{12}\rightarrow0$ limit at the smallest resolved separations (\secref{sec:results_data}); we nonetheless present the $z=0$ constraints for completeness, but they should be interpreted with caution for the same reason.

\begin{figure}
\includegraphics[width=\columnwidth]{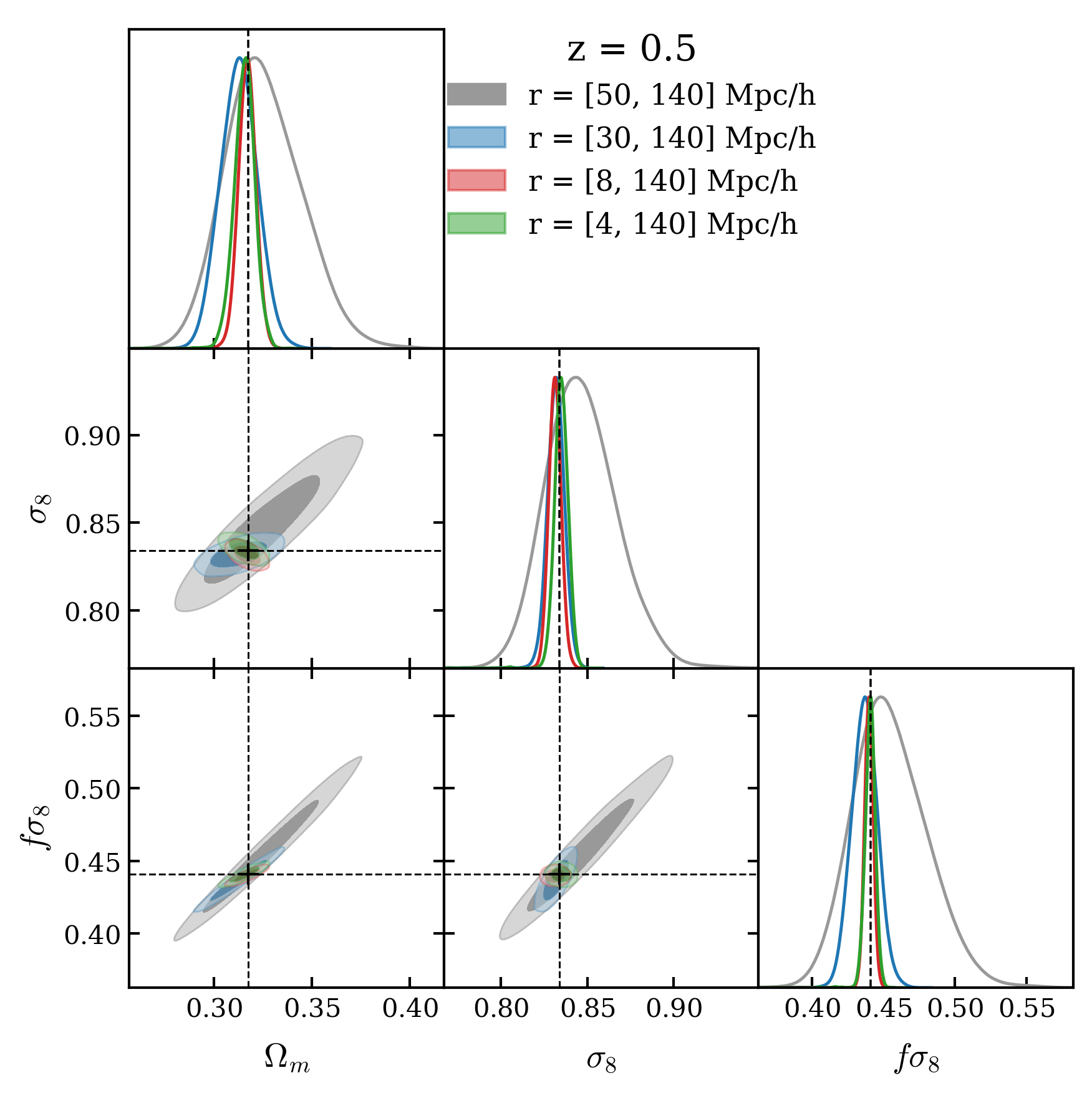}
\caption{\label{fig:trianglez05}  \textbf{Two-dimensional and one-dimensional marginalized posterior distributions} for $\sigma_8$, $\Omega_\mathrm{m}$, and $f\sigma_8$ at $z=0.5$, comparing four choices of minimum fitting scale, $r_{\rm min}\in\{50,30,8,4\}\,h^{-1}$Mpc (grey, blue, red, green), at fixed $r_{\rm max}=140\,h^{-1}$Mpc. Shaded regions show the $68\%$ and $95\%$ credible levels; dashed lines mark the fiducial \Quijote~ cosmology. The corresponding relative reduction in the $1\sigma$ uncertainty for each parameter is tabulated in Table~\ref{tab:rel_improvement} (Appendix~\ref{app:mcmc_stats}).}
\end{figure}

Figure~\ref{fig:trianglez05} shows the two-dimensional posterior distributions at $z=0.5$, comparing four choices of minimum fitting scale, $r_{\rm min}=50$, $30$, $8$, and $4\,h^{-1}$Mpc, at fixed $r_{\rm max}=140\,h^{-1}$Mpc. In all four cases the recovered best-fit values remain consistent with the fiducial \Quijote~ cosmology (dashed lines). We take $r_{\rm min}=50\,h^{-1}$Mpc as the baseline scale, since this marks the linear regime of streaming motions, where the full non-linear solution Eq. \eqref{eq:v12} agrees with the linear approximation Eq. \eqref{eq:v12lin} to within $5\%$ (\secref{sec:model}).

Progressively extending the fit to smaller, increasingly non-linear separations ($r_{\rm min}=30$, $8$, then $4\,h^{-1}$Mpc) tightens all four marginalized posteriors further. We adopt $r_{\rm min}=4\,h^{-1}$Mpc as the production scale at $z=0.5$, the smallest value supported by the residuals test (\secref{sec:results_data}).

We repeat the same analysis at $z=0$ in Fig.~\ref{fig:triange_z0} (Appendix~\ref{app:mcmc_stats}), restricting the discussion to $r_{\rm min}\geq8\,h^{-1}$Mpc: as at $z=0.5$, the recovered best-fit values remain consistent with the fiducial \Quijote~ cosmology, and tightening the fit from $r_{\rm min}=50$ to $8\,h^{-1}$Mpc improves all four marginalized posteriors. We do not adopt $r_{\rm min}=4\,h^{-1}$Mpc as a production scale at $z=0$: the residuals test (\secref{sec:results_data}) shows the model does not currently fit those bins well, so that contour is shown in Fig.~\ref{fig:triange_z0} for completeness only and should not be over-interpreted.

\begin{figure}
\includegraphics[width=\columnwidth]{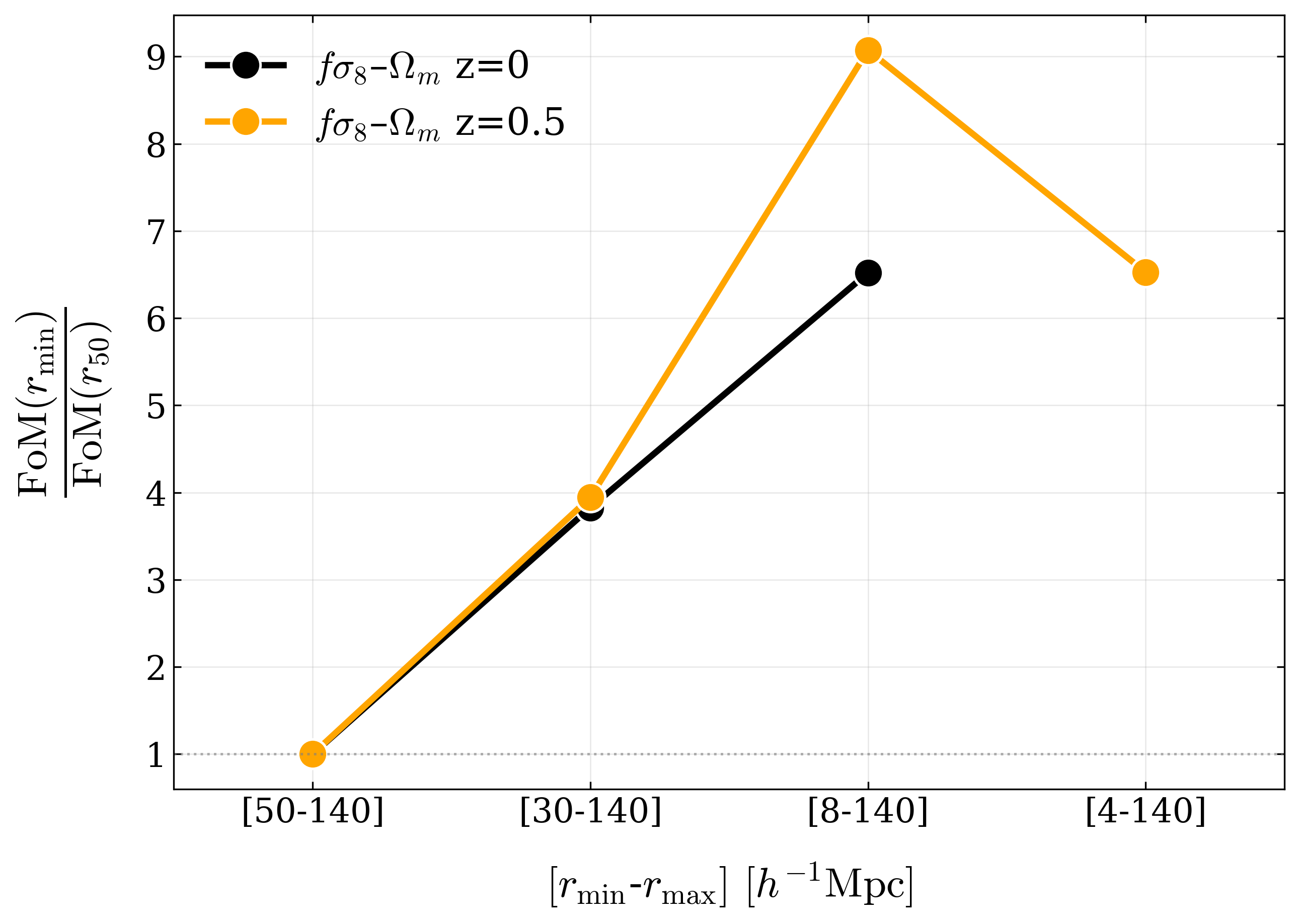}
\caption{\label{fig:fom} \textbf{Figure-of-Merit (FoM) in the $\Omega_\mathrm{m}$--$f\sigma_8$ plane} as a function of the minimum fitting scale $r_{\rm min}$, at fixed $r_{\rm max}=140\,h^{-1}$Mpc, normalised to its value at the $r_{\rm min}=50\,h^{-1}$Mpc baseline, for $z=0$ (black) and $z=0.5$ (orange). At the respective adopted production scales, the FoM ratio reaches $6.5\times$ at $r_{\rm min}=8\,h^{-1}$Mpc ($z=0$) and $6.5\times$ at $r_{\rm min}=4\,h^{-1}$Mpc ($z=0.5$), though the $z=0.5$ curve peaks higher still, at $9.1\times$, at the intermediate scale $r_{\rm min}=8\,h^{-1}$Mpc. The $z=0$, $r_{\rm min}=4\,h^{-1}$Mpc point is omitted, since that scale is not supported by the residuals test (\secref{sec:results_data}).}
\end{figure}

Figure~\ref{fig:fom} condenses the joint two-dimensional information visible in these contours into a single Figure-of-Merit, $\mathrm{FoM}\equiv1/\sqrt{\det Cov(\Omega_m, f\sigma_8)}$ (the inverse area of the $68\%$ credible ellipse) in the $\Omega_\mathrm{m}$--$f\sigma_8$ plane, normalised to its value at the $r_{\rm min}=50\,h^{-1}$Mpc baseline. At $z=0.5$, this FoM ratio reaches $3.9\times$ at $r_{\rm min}=30\,h^{-1}$Mpc, peaking at $9.1\times$ at $r_{\rm min}=8\,h^{-1}$Mpc before declining to $6.5\times$ at the adopted production scale $r_{\rm min}=4\,h^{-1}$Mpc (\secref{sec:results_data}); the decline towards smaller scales is related to the increasing model-data missmatch below  $r\simeq8$\Mpch (below $3\sigma$), where the model does not capture the pairwise velocity signal as well as it does over the $8$-$140$\Mpch range (below $1\sigma$) (see orange line in the bottom panel of Fig.~\ref{fig:v12model}); consequently, including the additional $4$-$8$\Mpch scales provides limited additional constraining power. This non-monotonic  behaviour mirrors the same pattern already noted for individual parameter uncertainties in Table~\ref{tab:rel_improvement} (Appendix~\ref{app:mcmc_stats}). At $z=0$, the FoM ratio reaches $3.8\times$ at $r_{\rm min}=30\,h^{-1}$Mpc, reaching $6.5\times$ at the adopted production scale $r_{\rm min}=8\,h^{-1}$Mpc; $r_{\rm min}=4\,h^{-1}$Mpc is not evaluated at $z=0$ (and is therefore omitted from Fig.~\ref{fig:fom}), since the $z=0$ model-data residuals exceed $3\sigma$ there (\secref{sec:results_data}) and the cut includes bins the model does not currently fit well.

To unpack which individual parameters drive this joint gain, we now turn to the marginal, one-dimensional constraints underlying the FoM, across scale and redshift.
Table~\ref{tab:mcmc_constraints} in Appendix~\ref{app:mcmc_stats} reports the full set of marginalized constraints at both redshifts. Among the adopted scales, $r_{\rm min}=8\,h^{-1}$Mpc gives the tightest $z=0$ constraints.
At $r_{\rm min}=50\,h^{-1}$Mpc, however, the $z=0$ constraints are systematically wider than their $z=0.5$ counterparts (e.g.\ $\sigma(f\sigma_8)=0.023$--$0.029$ at $z=0.5$ versus $0.025$--$0.033$ at $z=0$).

\begin{figure}
\includegraphics[width=\columnwidth]{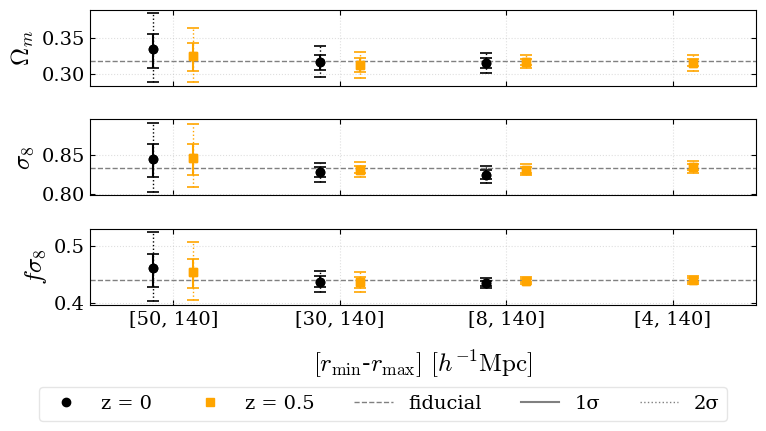}
\caption{\label{fig:scale} \textbf{Marginal constraints on $\Omega_\mathrm{m}$, $\sigma_8$, and $f\sigma_8$ as a function of the minimum fitting scale $r_{\rm min}$}, at fixed $r_{\rm max}=140\,h^{-1}$Mpc, for $z=0$ (black circles, $r_{\rm min}\in\{50,30,8\}\,h^{-1}$Mpc) and $z=0.5$ (orange squares, $r_{\rm min}\in\{50,30,8,4\}\,h^{-1}$Mpc). Solid and dotted error bars indicate the $68\%$ and $95\%$ equal-tailed intervals of the marginalised 1D posterior, respectively. Dashed horizontal lines mark the fiducial values of the simulations. Constraints are derived from MCMC chains with a burn-in fraction of $50\%$. As elsewhere in this section, $z=0$'s $r_{\rm min}=4\,h^{-1}$Mpc point is omitted, since that scale is not supported by the residuals test (\secref{sec:results_data}). }
\end{figure}

Figure~\ref{fig:scale} and Table~\ref{tab:rel_improvement} (Appendix~\ref{app:mcmc_stats}) quantify this marginal gain directly, taking $r_{\rm min}=50\,h^{-1}$Mpc as the baseline. At $z=0.5$, the $1\sigma$ uncertainty on $f\sigma_8$ improves by $64\%$ at $r_{\rm min}=30\,h^{-1}$Mpc and $88\%$ at $r_{\rm min}=8\,h^{-1}$Mpc, settling at $86\%$ at the adopted production scale $r_{\rm min}=4\,h^{-1}$Mpc. At $z=0$, the corresponding improvements are $67\%$ at $r_{\rm min}=30\,h^{-1}$Mpc, reaching $85\%$ at the adopted production scale $r_{\rm min}=8\,h^{-1}$Mpc; as elsewhere in this section, $r_{\rm min}=4\,h^{-1}$Mpc is not adopted at $z=0$ and is therefore omitted from Fig.~\ref{fig:scale}, for the reasons discussed above.

\section{\label{sec:discussion}Discussion}
\subsection{Resolution requirements and their limits}\label{sec:discussion_resolution}

\secref{sec:results_data} shows that \Quijote~ Mid-Resolution measurements fail to recover the expected \vot$\rightarrow0$ limit as $r\rightarrow0$ at $z=0$, unlike at $z=0.5$ and in TNG300-3. 
This likely reflects improper handling of particle velocities in the \Quijote~ boxes at that resolution and redshift, rather than a failure of the model, since the same pipeline recovers the limit in the other cases. Both the full ($100\%$) Mid-Resolution snapshot and the independent High-Resolution configuration show the same failure at $z=0$ (\figref{fig:v12model}), ruling out particle thinning and resolution as the cause. The tests presented here do not uniquely identify the origin of this discrepancy; we therefore conservatively exclude the affected scales from the production inference.
Despite this, \Quijote~ is still the better primary suite for two reasons. First, it yields substantially better overall model-data agreement than \TNG: \figref{fig:halofit} shows \Quijote~ ratios remain within or near $\pm5\%$ across most scales, while \TNG~ deviates by more than $10\%$ over an extended range (\secref{sec:results_validation}). Second, \TNG's smaller box ($L_{\rm box}=205\,\Mpch$) cannot reliably support the $r_{\rm max}=140\,\Mpch$ pair separations used in our production analysis, since $140\,\Mpch$ exceeds half its box length, whereas \Quijote's larger volume ($L_{\rm box}=1000\,\Mpch$) can (\appref{app:lbox}).

More broadly, the $r_{\rm min}$ values adopted in \secref{sec:constraints} are set by comparing a noiseless, dark-matter-only $N$-body prediction to a clean simulation measurement. Real spectroscopic surveys introduce additional small-scale systematics that this comparison does not capture such as fibre collisions, redshift measurement errors, and the velocity bias between galaxies (or halos) and the underlying dark matter, among others. The scale cuts derived here therefore represent a lower bound set by simulation resolution, not a complete prescription for applying \pairvel~ to real data; a full observational analysis would need to establish its own $r_{\rm min}$ accounting for these additional effects.

We also note that the TNG300-3 cross-check used throughout \secref{sec:results_data} and \secref{sec:results_validation} relies on a single realisation, with no associated covariance estimate; its role here is qualitative, confirming that the $\text{v}_{12}\rightarrow0$ behaviour is achievable at higher resolution, rather than providing an independent quantitative validation of the production pipeline. A higher-resolution, multi-realisation simulation suite,  extending the \Quijote~ approach to smaller particle masses at fixed volume, would allow this cross-check to be placed on the same statistical footing as the rest of the analysis.

\subsection{Validating the non-linear prescription}\label{sec:discussion_validation}

The core of this pairwise velocities model is rooted in the time evolution of the two point clustering signal. Predicting this evolution accurately requires two distinct ingredients built from the non-linear power spectrum: the correlation function $\xi(r,a)$ itself, and its scale-factor derivative $\partial_a\bar{\xi}(r,a)$, which carries the growth signal. It is therefore crucial to test the best choice for both. In our case, we show that an improper choice of the first ingredient -- the Takahashi non-linear prescription -- leads to a sensitivity in the latter, i.e.\ the time derivative, to the discretization of the redshift array used to evaluate it (\secref{sec:results_validation}), giving spurious deviations in the model that could erroneously be interpreted as cosmological signatures. \code{HMcode-2020} does not show this sensitivity.

\subsection{\vot~ and cosmological degeneracies}\label{sec:discussion_degeneracy}

The growth rate $f$ is not sampled as an independent parameter in our MCMC: it is computed numerically at each sampled cosmology from the \camb~growth history, $f(a)=d\ln D/d\ln a$ via central differences, and is therefore fully determined by $\Omega_\mathrm{m}$ (and the other sampled parameters) rather than free to vary independently.
Consequently, our result should not be read as breaking the classic $f$--$\sigma_8$ degeneracy familiar from redshift-space distortion and peculiar-velocity analyses, which requires measuring $f$ independently of a bias-$\sigma_8$ combination. What we demonstrate instead is that \pairvel separately constrains
$\Omega_m$ and $\sigma_8$ through their distinct scale-dependent imprints on the pairwise velocity field (\secref{sec:sensitivity}), yielding complementary constraints on the growth history within the assumed $\Lambda$CDM parameter space.

The $\Omega_\mathrm{m}$ response identified in \secref{sec:sensitivity} sets \pairvel~apart from density-only clustering statistics: $\partial\ln \text{v}_{12}/\partial\ln\Omega_\mathrm{m}$ changes sign near $r\approx100$--$110\,\;\text{Mpc}$, growing positive at intermediate scales, where the shape of the matter transfer function dominates, and turning negative at larger separations, where the enhanced expansion rate suppresses the normalised infall signal. This sign reversal is not shown by $h,\;\sigma_8$, and it is absent from $\xi(r,a)$ altogether. Because this feature has no counterpart in the equal-time two-point correlation function, it provides a distinct scale-dependent cosmological sensitivity, and is one mechanism contributing to the separation of $\Omega_m$ and $\sigma_8$ in the \vot\ posterior.

We verified this expectation directly by re-running the production chains at both adopted configurations with $h$ sampled freely rather than fixed. At $z=0.5$ ($r_{\rm min}=4\,\Mpch$), the marginalized $h$ posterior lands at $0.818\pm0.026$, $5.7\sigma$ from the \Quijote\ fiducial value $h=0.6711$, and drags $\Omega_\mathrm{m}$ and $\sigma_8$ to $0.216\pm0.013$ and $0.931\pm0.017$ respectively -- $2.7\times$ and $4.1\times$ wider than, and tens of $\sigma$ offset from, the $h$-fixed constraints of \tabref{tab:mcmc_constraints}. The $z=0$ configuration ($r_{\rm min}=8\,\Mpch$) shows the same behaviour: $h=0.853\pm0.030$ ($6.0\sigma$ from fiducial), with $\Omega_\mathrm{m}$ and $\sigma_8$ similarly displaced. Rather than merely broadening the posterior around the true cosmology, an unconstrained $h$ moves the fit to a different, self-consistent point on the degeneracy ridge, confirming that \pairvel alone cannot break this degeneracy at the scales used here, and justifying our choice to fix $h$ for the production constraints.

If one wishes to constrain simultaneously the amount of matter, the clustering amplitude, the growth rate, and the Hubble constant, access to separations of order $1\,\mathrm{Mpc}$ (approximately $0.7\,\Mpch$ at the \Quijote~fiducial cosmology) is required (\secref{sec:sensitivity}), even in this simplified, dark-matter-only idealised scenario. This demonstrates the importance of pushing towards better modelling in the non-linear regime for peculiar velocities.

\subsection{Implications for current and future velocity surveys}\label{sec:discussion_surveys}

At the adopted production scales, in our idealised gravity-only simulation analysis, the marginalized constraints reach $\sigma(\Omega_\mathrm{m})/\Omega_\mathrm{m}\approx1.6\%$ and $\sigma(\sigma_8)/\sigma_8\approx0.5\%$ at $z=0.5$ ($r_{\rm min}=4\,\Mpch$), against $2.2\%$ and $0.7\%$, respectively, at $z=0$ ($r_{\rm min}=8\,\Mpch$; \tabref{tab:mcmc_constraints}). 
The corresponding derived growth constraints, $f\sigma_8=0.441^{+0.004}_{-0.003}$ at $z=0.5$ and $f\sigma_8=0.435^{+0.005}_{-0.004}$ at $z=0$, are broadly consistent with the range of low-redshift $f\sigma_8$ measurements compiled from existing peculiar-velocity surveys by \citet{turnerDESIDR1Peculiar2025}. 
Reaching this precision from \pairvel~alone required resolving pair separations down to $4$--$8\,\Mpch$ (\secref{sec:results_data}), well inside the $\gtrsim20$--$30\,\Mpch$ linear/quasi-linear regime to which current kSZ pairwise analyses restrict their modelling \citep{hadzhiyskaProbingCosmicVelocities2025, gongDetectionPairwiseKinematic2025}.

Exploiting these non-linear scales observationally requires reliable measurements and modelling of pairwise motions well below the quasi-linear regime. For spectroscopic and peculiar-velocity samples this requires controlling tracer selection, sparse sampling, redshift and distance errors, fibre assignment, and the relation between galaxy/halo and matter pairwise velocities; kSZ applications additionally require modelling the connection between the measured temperature signal, halo gas/optical depth, and the underlying pairwise velocity. but the constraining power demonstrated here indicates the potential pay-off is substantial. This motivates the development of simulation-calibrated non-linear pairwise-velocity models as an ingredient for future analyses across distinct observational channels: direct peculiar-velocity surveys such as 4MOST/4HS \citep{dejong4MOST4metreMultiobject2012,taylor4MOSTHemisphereSurvey2023}, redshift-space clustering measurements from surveys such as DESI, Euclid, and PFS, and kSZ detections from CMB experiments.

\subsection{Limitations and future work}\label{sec:discussion_limitations}

Our validation strategy targets \LCDM~directly, at the resolution afforded by the \Quijote~and \TNGtt~suites (\secref{sec:results_validation}), which complements the high-resolution, scale-free (self-similar) convergence tests of \citet{maleubreConstrainingAccuracyPairwise2023}; the two approaches trade off cosmological realism against attainable resolution, and agreement between them would strengthen confidence in the small-scale limit of \pairvel~in either framework. 
We further verified that our results are insensitive to the box-size fundamental mode for the volumes considered (\appref{app:lbox}), though this should be revisited for the smaller effective volumes typical of realistic survey footprints or mocks.

Our analysis uses dark-matter-only pairwise velocities measured directly from simulations, without a galaxy bias or redshift-space distortion model. 
\citet{kuruvillaImprintBaryonsMassive2020} show that, in several state-of-the-art hydrodynamical suites (Illustris-TNG, EAGLE, BAHAMAS, and cosmo-OWLS), the mean pairwise velocity of the total matter differs from the corresponding collisionless simulation by $\leq1$--$3\%$ at separations of a few Mpc; combined with the well-established result that baryons couple more weakly to the velocity field than to the density field \citep{hellwingEffectBaryonsRedshift2016}, this suggests that baryonic corrections to the \emph{matter} pairwise velocity are at the percent level on the scales used in this work ($\gtrsim4$--$8\,\Mpch$). 
This does not, however, remove the separate problem of relating the matter statistic to the pairwise velocities of a selected galaxy or halo population.
A further complication is that the exact pair-conservation equation employed here applies directly to a conserved particle population, whereas halo and galaxy populations evolve through formation, merging, and sample selection. A realistic tracer implementation therefore requires testing how these effects modify the connection between the tracer correlation function and its mean pairwise velocity.
Extending the likelihood to biased tracers in redshift space, and validating against realistic mock catalogues with survey-like selection functions, is the necessary next step toward applying this framework to observational data; the specific systematics involved differ by observational channel. Direct peculiar-velocity catalogues introduce distance-indicator errors, radial selection effects, and observer-dependent systematics, whereas kSZ applications additionally require modelling the relation between the observed temperature signal and the halo gas/optical depth. We treat these as distinct future applications rather than a single observational pipeline.

\section{\label{sec:conclusions}Conclusions}
We have constructed and tested a cosmological likelihood for the mean pairwise velocity \pairvel, built on the exact pair conservation equation and the non-linear modelling framework developed in \citet{jaberDynamicsPairwiseMotions2023}. 
Across the \Quijote~and \TNGtt~validation suite, we tested the sensitivity of the inference framework to particle thinning, simulation resolution, box size, and the adopted non-linear power-spectrum prescription, with \code{HMcode-2020} adopted as the fiducial ingredient after providing the most stable predictions among the prescriptions tested for the scale-factor derivative $\partial_a\bar{\xi}(r,a)$ (\secref{sec:results_validation}). 
Our resolution and convergence tests complement the convergence tests of \citet{maleubreConstrainingAccuracyPairwise2023}, who characterise the accuracy of \pairvel~in $N$-body simulations at the percent level using very high-resolution scale-free (self-similar) cosmologies; our validation instead targets the \LCDM~case directly, at the resolution afforded by the \Quijote~and \TNGtt~suites, trading their idealised scale-invariance for a direct test in a $\Lambda$CDM simulation setup.

Using this validated likelihood, in an idealised gravity-matter-only analysis of the \Quijote~simulations, we obtain percent-level constraints on $\Omega_\mathrm{m}$ and $\sigma_8$ ($h$ fixed at its \Quijote~fiducial value, \secref{sec:sensitivity}), and their derived counterpart $f\sigma_8$, from \Quijote~Mid-Resolution 10\% production runs. Extending the fit to smaller, non-linear separations beyond the linear-regime baseline substantially improves these constraints: at the respective adopted production scales, $r_{\rm min}=4\,\Mpch$ at $z=0.5$ and $r_{\rm min}=8\,\Mpch$ at $z=0$ (\secref{sec:results_data}), the $1\sigma$ uncertainty on $\Omega_\mathrm{m}$ and $\sigma_8$ tightens by up to $74\%$ and $81\%$ at $z=0.5$ and $70\%$ and $74\%$ at $z=0$, and $\sigma(f\sigma_8)$ tightens by up to $86\%$, and the $\Omega_\mathrm{m}$--$f\sigma_8$ Figure-of-Merit ratio reaches $6.5\times$ at both adopted production scales (peaking higher still, at $9.1\times$, at $z=0.5$'s intermediate $r_{\rm min}=8\,\Mpch$ cut before declining), all relative to the $r_{\rm min}=50\,\Mpch$ baseline (\secref{sec:constraints}), demonstrating that these scales carry substantial cosmological information within the idealised inference setup considered here.

The natural next step is to extend this framework beyond gravity-only pairwise velocities: testing the pair-conservation framework for evolving halo and galaxy populations, together with tracer selection, redshift-space effects, and realistic survey systematics, is required before applying the method to observed peculiar-velocity, spectroscopic, or kSZ data \citep{hadzhiyskaProbingCosmicVelocities2025, gongDetectionPairwiseKinematic2025}. Beyond a stand-alone \pairvel~likelihood, combining it with distance measurements from Type~Ia supernovae \citep{broutPantheonAnalysisCosmological2022}, baryon acoustic oscillations and redshift-space distortions \citep{desicollaborationDESIDR2Results2025}, weak lensing, and the CMB \citep{aghanimPlanck2018Results2018} offers a route to joint, multi-probe constraints on the growth of structure and expansion history. The scale-dependent $\Omega_\mathrm{m}$ response identified in \secref{sec:sensitivity}, whose sign reversal has no counterpart in the two-point correlation function (\secref{sec:discussion_degeneracy}), indicates that \pairvel~carries cosmological information not already encoded in density-only clustering statistics. More generally, simultaneously constraining the amount of matter, the clustering amplitude, the growth rate, and the Hubble constant with \pairvel~alone would require access to separations of order $1\,\mathrm{Mpc}$ (approximately $0.7\,\Mpch$), even within this idealised, dark-matter-only scenario; achieving this reach motivates continued development of accurate non-linear pairwise-velocity modelling at these scales. Quantifying the amount of genuinely independent information a joint analysis could extract, however, requires the cross-covariances between \pairvel~and these density-based probes to be modelled explicitly, which we defer to future work.
We plan to pursue this bias extension and multi-probe combination in follow-up work.

\appendix

\section{Robustness of the Theoretical Modelling to Numerical Artifacts}

This Appendix presents a detailed examination of the numerical stability of our model/theoretical framework, where the validation tests originally conducted for Quijote in the main text are here extended to the TNG300-3 volume.
We find that the model performance is consistent across both suites, confirming that our results are robust against these numerical choices regardless of the simulation architecture.

\subsection{Dependence of the \vot~ model to the box-size fundamental mode.}
\label{app:lbox}

\begin{figure}
\includegraphics[width=0.99\columnwidth]{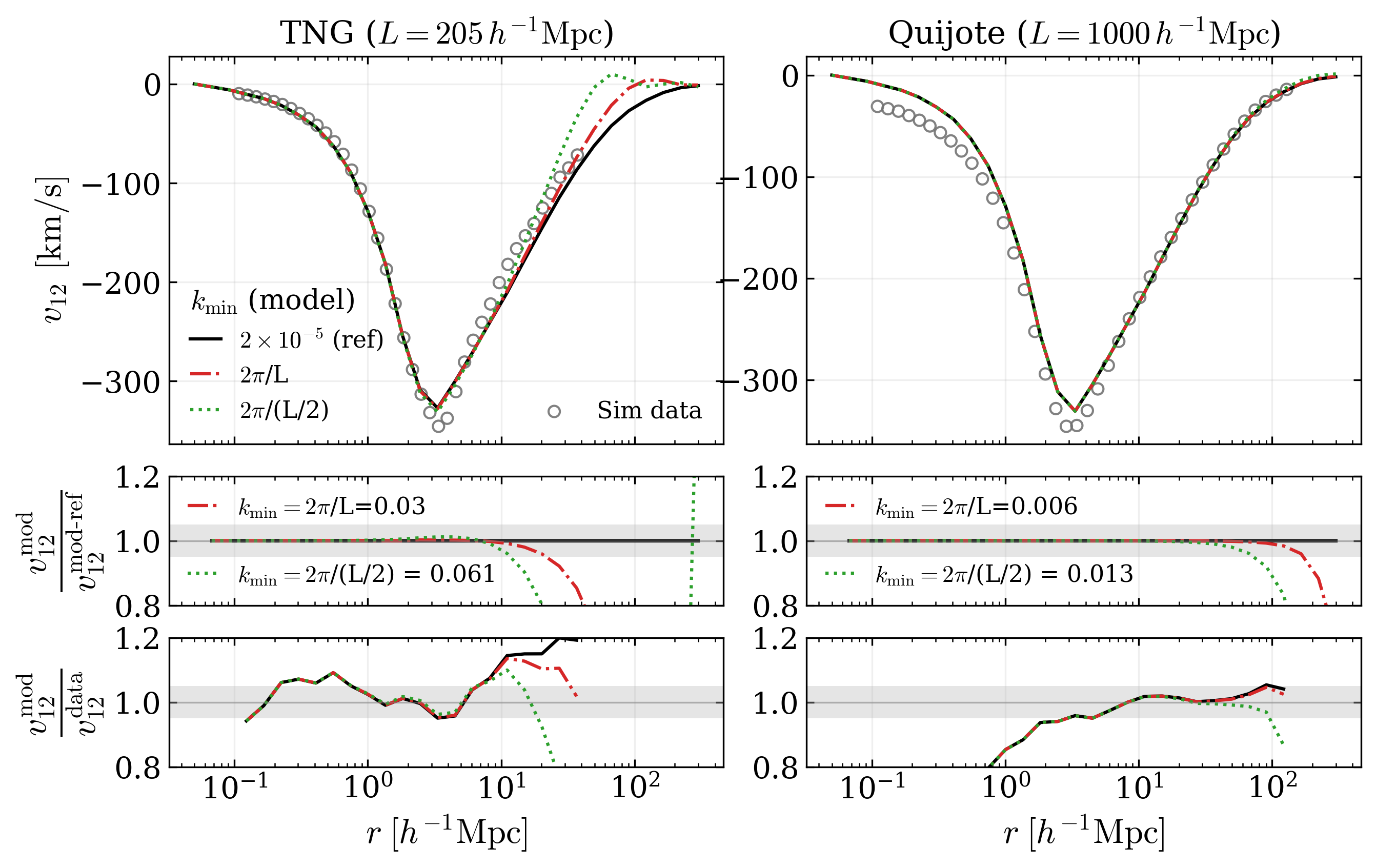}
\caption{\label{fig:v12_tng_kmin} Sensitivity of the mean pairwise velocity $\text{v}_{12}(r)$ to the fundamental mode of the simulation volume ($k_f = 2\pi/L_{box}$). \textbf{Left panels} correspond to the \TNG~ simulation ($L_{box} = 205$ Mpc/h), while \textbf{right panels} correspond to the Quijote ensemble ($L_{box} = 1000$ Mpc/h). \textbf{The top row} presents a comparison between simulation measurements (grey points) at $z=0$ and theoretical predictions obtained by integrating the power spectrum $P(k)$ from different minimum wave numbers: the fiducial choice $k_{min} = 2 \times 10^{-5}$ h/Mpc (solid black), the box fundamental mode $k_f$ (dash-dotted red), and $2k_f$ (dotted green). \textbf{The middle and bottom rows} display, respectively, the ratios of the models relative to the fiducial case and to the simulation measurements. At large separations, the predicted $\text{v}_{12}$ exhibits an increasing dependence on the adopted $k_{min}$, which reflects the contribution of large-scale modes that cannot be fully sampled within the finite simulation volume.}
\end{figure}

To assess the impact of finite simulation volume on pairwise motions, we compare the modelled \pairvel~obtained by truncating the power spectrum at the fundamental mode $k_f$ associated with each box. 
As shown in Fig.~\ref{fig:v12_tng_kmin}, the \pairvel~predictions at large separations are highly sensitive to the absence of long-wavelength modes, consistent with the general sensitivity of velocity statistics to missing large-scale power \citep{scoccimarroRedshiftspaceDistortionsPairwise2004}.
For \TNG~, the relatively small box size ($L_{box} = 205$ Mpc/h) implies a relatively high fundamental wavenumber ($k_f \approx 0.03$ h/Mpc), leading to a noticeable deviation from the infinite-volume default prediction. 
By contrast, the larger \Quijote~ volume ($L_{box} = 1000$ Mpc/h) has a much smaller $k_f \approx 0.006$ h/Mpc, so the truncated model remains close to the default prediction over the relevant separations, indicating that the long-wavelength modes needed to model the large-scale velocity field are sufficiently captured for our analysis.
This comparison validates our use of \Quijote~ for the main analysis, as it ensures that the power missing due to the box size does not bias the inferred velocity statistics at the scales of interest, while \TNG~ serves as a benchmark for detecting these finite-volume effects.

\section{Covariance Matrix}

We estimate the bin-to-bin covariance matrix from $N=100$ independent realizations using Eq. \ref{eq:covmat}. We verified the numerical convergence of the covariance estimate by progressively increasing the number of realizations used in the calculation. The correlation structure and covariance amplitudes were found to stabilize once approximately 50 realizations were included, with only negligible changes observed thereafter. We therefore adopt $N=100$ realizations for the fiducial covariance estimate, providing a conservative margin beyond the convergence threshold while remaining computationally efficient. This convergence test establishes stability of the covariance elements themselves, but does not by itself guarantee convergence of $C^{-1}$ or of the resulting posterior widths, since matrix inversion can amplify residual finite-sample noise even where individual elements have already stabilized. The simple Hartlap debiasing rescaling \cite{Hartlap2007},
$\alpha=(N-p-2)/(N-1)$, ranges from $\alpha\approx0.79$ at $p=20$ ($r_{\rm min}=4\,h^{-1}$Mpc) to $\alpha\approx0.93$ at $p=6$ ($r_{\rm min}=50\,h^{-1}$Mpc) for the bin counts used here, corresponding to a $4$--$13\%$ underestimate of parameter uncertainties if the naive covariance were used uncorrected; since this only corrects the expectation value of $C^{-1}$ and not its own sampling uncertainty, we found it insufficient on its own. 
We address this directly by adopting the multivariate $t$-distribution likelihood of \citet{sellentinParameterInferenceEstimated2016} (\secref{sec:likelihood}), which analytically marginalises over the sampling uncertainty of $C$ itself rather than relying on the point estimate $C^{-1}$.

We additionally verified the Gaussianity assumption underlying the Sellentin \& Heavens likelihood by computing the skewness and kurtosis of \pairvel~across the available realizations, independently for each radial bin at the adopted production scales. After a Bonferroni correction for the number of bins tested, no bin shows a statistically significant deviation from Gaussianity at either redshift, consistent with the central-limit-theorem averaging implicit in \pairvel~despite individual pair velocities being expected to develop skewed, non-Gaussian tails under non-linear evolution \citep{juszkiewiczSkewedExponentialPairwise1998}.

\section{Full MCMC Constraints}
\label{app:mcmc_stats}

\begin{figure}
\includegraphics[width=\columnwidth]{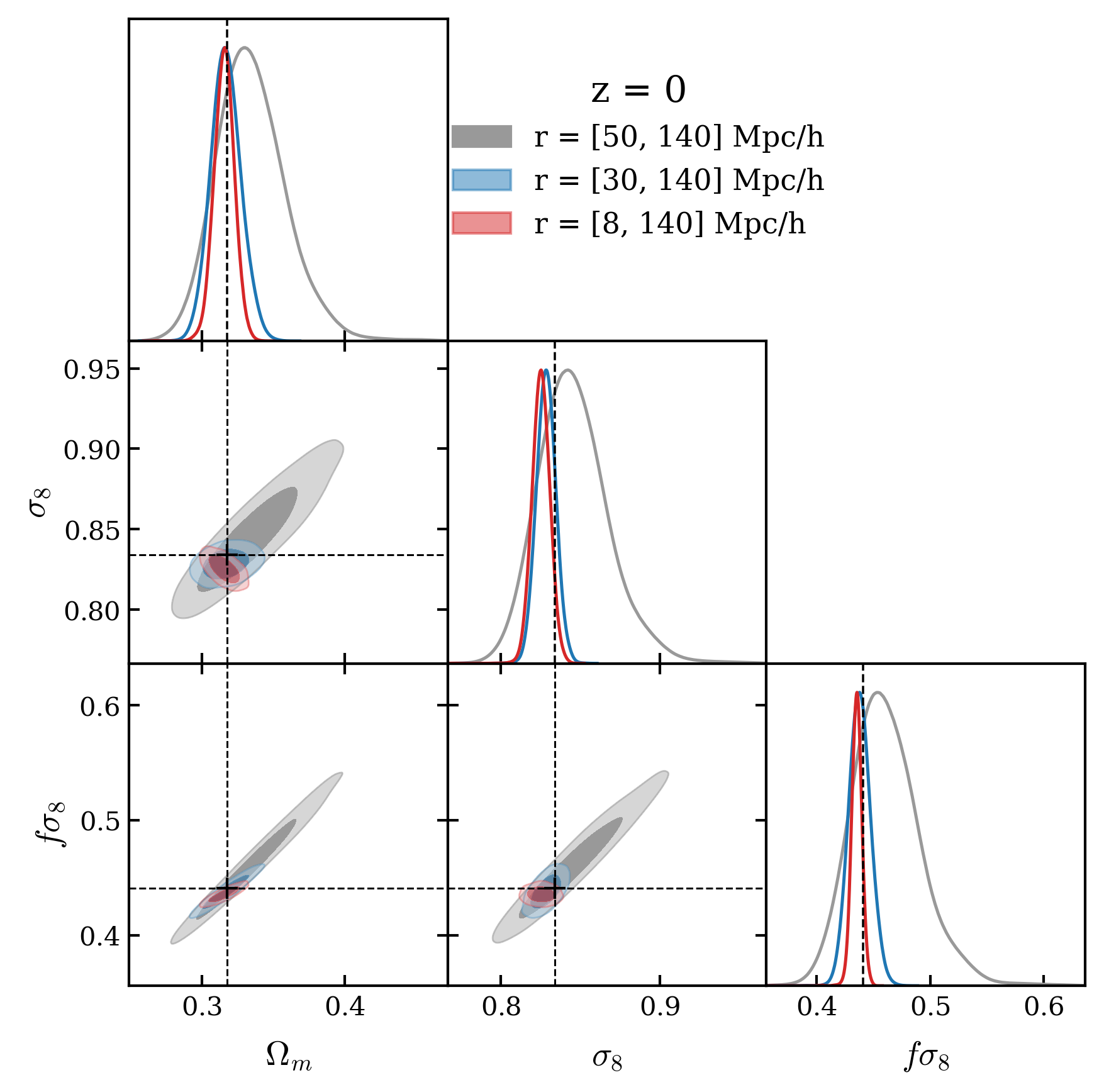}
\caption{\label{fig:triange_z0} \textbf{Two-dimensional and one-dimensional marginalized posterior distributions for $\sigma_8$, $\Omega_\mathrm{m}$, and $f\sigma_8$ at $z=0$}, comparing three choices of minimum fitting scale, $r_{\rm min}\in\{50,30,8\}\,h^{-1}$Mpc, at fixed $r_{\rm max}=140\,h^{-1}$Mpc. Shaded regions show the $68\%$ and $95\%$ credible levels; dashed lines mark the fiducial \Quijote~ cosmology. Unlike at $z=0.5$ (\figref{fig:trianglez05}), $r_{\rm min}=4\,h^{-1}$Mpc is omitted here, since that scale is not supported by the residuals test (\secref{sec:results_data}); the main text restricts its discussion to $r_{\rm min}\geq8\,h^{-1}$Mpc accordingly.}
\end{figure}

This Appendix collects the full set of marginalized posterior constraints supporting \secref{sec:constraints}. Fig.~\ref{fig:triange_z0} shows the $z=0$ triangle plot referenced there; Table~\ref{tab:mcmc_constraints} reports the complete marginalized constraints, including the directly sampled parameters, at both redshifts and all scale cuts; Table~\ref{tab:rel_improvement} tabulates the corresponding relative improvement in each parameter's $1\sigma$ uncertainty with respect to the $r_{\rm min}=50\,h^{-1}$Mpc baseline.

\begin{table*}
\caption{Marginalized parameter constraints (median with 68\% credible intervals) at $z=0$ and $z=0.5$ for different minimum pair separation scales.}
\label{tab:mcmc_constraints}
\begin{tabular}{l|ccc|cccc}
\toprule
 & \multicolumn{3}{c}{$z = 0$} & \multicolumn{4}{c}{$z = 0.5$} \\
\cmidrule(lr){2-4} \cmidrule(lr){5-8}
Parameter & $r_\mathrm{min} = 8$ & $r_\mathrm{min} = 30$ & $r_\mathrm{min} = 50$ & $r_\mathrm{min} = 4$ & $r_\mathrm{min} = 8$ & $r_\mathrm{min} = 30$ & $r_\mathrm{min} = 50$ \\[6pt]
\midrule
$\log(10^{10}A_s)$ & $3.049^{+0.040}_{-0.044}$ & $3.052^{+0.047}_{-0.047}$ & $3.018^{+0.059}_{-0.059}$ & $3.072^{+0.028}_{-0.031}$ & $3.057^{+0.026}_{-0.026}$ & $3.078^{+0.038}_{-0.041}$ & $3.058^{+0.045}_{-0.050}$ \\[6pt]
$\Omega_c h^2$ & $0.120^{+0.003}_{-0.003}$ & $0.121^{+0.005}_{-0.005}$ & $0.129^{+0.009}_{-0.012}$ & $0.120^{+0.002}_{-0.002}$ & $0.121^{+0.002}_{-0.002}$ & $0.119^{+0.004}_{-0.004}$ & $0.125^{+0.008}_{-0.010}$ \\[6pt]
$\sigma_8$ & $0.825^{+0.006}_{-0.005}$ & $0.828^{+0.006}_{-0.006}$ & $0.846^{+0.019}_{-0.024}$ & $0.835^{+0.004}_{-0.004}$ & $0.831^{+0.004}_{-0.003}$ & $0.832^{+0.005}_{-0.005}$ & $0.847^{+0.018}_{-0.022}$ \\[6pt]
$\Omega_m$ & $0.316^{+0.007}_{-0.007}$ & $0.317^{+0.010}_{-0.011}$ & $0.335^{+0.021}_{-0.026}$ & $0.316^{+0.005}_{-0.005}$ & $0.317^{+0.004}_{-0.004}$ & $0.313^{+0.009}_{-0.010}$ & $0.326^{+0.017}_{-0.022}$ \\[6pt]
$f\sigma_8$ & $0.435^{+0.005}_{-0.004}$ & $0.438^{+0.010}_{-0.010}$ & $0.461^{+0.025}_{-0.033}$ & $0.441^{+0.004}_{-0.003}$ & $0.440^{+0.003}_{-0.003}$ & $0.437^{+0.009}_{-0.009}$ & $0.455^{+0.023}_{-0.029}$ \\[6pt]
\bottomrule
\end{tabular}
\end{table*}

\begin{table}
\centering
\caption{Relative improvement in parameter uncertainties with respect to $r_\mathrm{min} = 50\,h^{-1}\,\mathrm{Mpc}$.}
\label{tab:rel_improvement}
\resizebox{\ifdim\width>\columnwidth\columnwidth\else\width\fi}{!}{%
\begin{tabular}{lrrrrr}
\toprule
 & $\log(10^{10}A_s)$ & $\Omega_c h^2$ & $\sigma_8$ & $\Omega_m$ & $f\sigma_8$ \\
Scale & (\%) & (\%) & (\%) & (\%) & (\%) \\
\midrule
\multicolumn{6}{l}{\textit{z = 0}} \\
\midrule
$r_{\rm min} = 8$ & $28.7$ & $70.4$ & $73.8$ & $70.4$ & $84.7$ \\
$r_{\rm min} = 30$ & $20.1$ & $55.3$ & $71.6$ & $55.3$ & $67.0$ \\
\midrule
\multicolumn{6}{l}{\textit{z = 0.5}} \\
\midrule
$r_{\rm min} = 4$ & $37.5$ & $73.7$ & $81.0$ & $73.7$ & $86.4$ \\
$r_{\rm min} = 8$ & $46.0$ & $77.3$ & $82.5$ & $77.3$ & $88.3$\\
$r_{\rm min} = 30$ & $17.1$ & $51.5$ & $76.2$ & $51.5$ & $64.3$ \\
\bottomrule
\end{tabular}%
}
\end{table}


\begin{acknowledgments}

The authors express their gratitude for the support received from the PAIRS project via the SONATA grant no.\ 2023/51/D/ST9/02919.

We gratefully acknowledge Polish high-performance computing infrastructure PLGrid (HPC Center: CI TASK) for providing computer facilities and support within computational grant no. PLG/2025/019011. We are grateful to the team maintaining the Kepler server, gracefully hosted at the Nicolaus Copernicus Astronomical Centre (CAMK) in Warsaw.

\textit{Software}: This work used \texttt{Halotools}, part of the \astropy{} project \citep{the_astropy_collaboration_astropy_2013,the_astropy_collaboration_astropy_2018}, \matplotlib{} \citep{hunter_matplotlib_2007}, \numpy{} \citep{walt_numpy_2011,harris_array_2020}, \python{} \citep{van_rossum_python_2009}, \scipy{} \citep{jones_scipy_2011,virtanen_scipy_2020}, \camb~and \pycamb, \code{Cobaya} \citep{torradoCobayaCodeBayesian2021}, and \texttt{GetDist}.

This research also made use of the NASA Astrophysics Data System~(\url{http://adsabs.harvard.edu/}) and the arXiv e-print service (\url{http://arxiv.org/}). We thank their developers for maintaining them and making them freely available.

\end{acknowledgments}

\newpage

\bibliography{bibs/v12_mcmc_refs_for_paper_group}

\end{document}